\documentclass[a4paper,12pt]{article}
\usepackage{amssymb}
\begin{document}
\begin{center}
{\large\bf The Unitary Mechanism of Infrared Freezing in \\
        QCD with massive gluons}\\
\vspace{4mm}

{\large D.V.Shirkov }

\vspace{0.2cm}
{\it Bogoliubov Theor. Lab, JINR, Dubna, 141980, Russia}\\
{\it shirkovd@thsun1.jinr.ru}
\end{center}
\begin{abstract}
{\footnotesize A ``natural" model for the QCD invariant (running) coupling,
free of the IR singularity, is proposed. It is based upon the hypothesis of
finite gluon mass $\,m_{\rm gl}$ existence and, technically, uses an accurate
treating of threshold behavior of Feynman diagram contribution. The model
 correlates with the unitarity condition.

\indent  Quantitative estimate, performed in the one-loop approximation
yields a reasonable lower bound for this mass $m_{\rm gl}>150$ MeV
and a smooth IR freezing at the level  $\alpha_s(Q^2)\lesssim 1$.}
\end{abstract}

\section{Introduction}

The issue of the infrared (IR) behavior of the strong interaction becomes
more and more actual from the physical point of view along with the further
experimental data accumulation. In the perturbative quantum chromodynamics
(pQCD) this behavior is burdened with ``unphysical" singularities marked
with the so called scale parameter $\,\Lambda \simeq 300\,$ MeV.  These
singularities contradict some general principles of the local QTF. In the
``small momentum transfer region" $\,Q \lesssim 3\Lambda\,$ they violate the
weak coupling regime and complicate theoretical interpretation of data.

   Quite recently attempts have been made to devise models for invariant
(running) coupling $\,\alpha_s(Q^2)\,$ and for observables free of
singularities in the IR region. One idea was to exploit the scheme
arbitrariness of the third beta-function coefficient $\,\beta_3$. Performing
an ``optimization" of the beta-function the authors of paper \cite{ms} argued
for special solution with an IR fixed point and $\,\alpha_s(Q^2)< 1\,$
bounded in the whole region $\,Q^2\leq 1\/ {\rm GeV}^2\,$.  In the paper
\cite{sim} the allowance was made for nonperturbative contributions into
$\,\alpha_s(Q^2)\,$ emerged from the vacuum QCD background fields. Here, the
adjusting parameter, the background (hybrid) gluon mass $\,M_B$, enters as an
IR regulator $\,\ln Q^2\to\ln (Q^2+M_B^2)$ of the gluonic logs.  At
$\,M_B=1.5\,$ GeV the running coupling maximum value turned out to be
$\,\alpha_s(0)\simeq 0.4\,$.  One more trick \cite{aa96-97} uses an
  imperative of the K\"all\'en--Lehmann analyticity in the complex $\,Q^2\,$
plane. Effectively, it results in the smooth freezing of $\,\alpha_s(Q^2)\,$
at the level of 0.5 -- 0.7 for 0.1\ ${\rm GeV}^2 \leq Q^2 \leq 1 \ {\rm
GeV}^2\,$ and in the universal, i.e., loop--independent IR limiting value
 $\,\alpha_s^{max}=\alpha_s(0)=4\pi/\beta_1\simeq 1.4\,$.

 In the last, so called ``invariant analytic", approach (for a recent review,
see \cite{tmp99}) the finiteness and smoothness of the $\,\alpha_s(Q^2)\,$
behavior in the IR region is achieved only due to the analyticity imperative,
without introducing any adjustable parameters. Qualitatively, the effect of
smooth freezing arises here due to the addition of a power (in the $\,Q^2\,$
variable) and nonanalytic (in the coupling constant $\,\alpha_s$), i.e.,
nonperturbative terms which restore the K\"all\'en--Lehmann analyticity for
renormalization -- invariant quantities.

 In this note, we consider one more possibility of constructing the invariant
coupling $\,\alpha_s(Q^2)\,$ that is free of unphysical singularities and
{\it does not involve explicit nonperturbative contributions}. Here, the
$\,Q^2\,$ power contributions appear due to threshold effects and an essential
physical ingredient is the assumption of the {\sf finite gluon mass
existence}. \par

 Our model expression for $\,\alpha_s(Q^2)\,$ is obtained by the
renormalization group (RG) summation of mass--dependent one--loop Feynman
diagrams contribution --- see, below, Eqs. (\ref{a1rg}) and (\ref{A1mass}). It
depends upon the gluonic $\,m_{\rm gl}\,$ and light quark $\,m_{\rm u,d,s}\,$
masses, in the IR region $Q^2>0\,$ obeys a nonsingular behavior with a finite
limiting value $\,\alpha_s(0)\,$, and as $\,Q^2/m^2 \to \infty \,$
smoothly transits into the usual asymptotic freedom formula.

\section{Massive loops}

  Our starting point is quite simple and natural --- we propose to take into
the account the threshold mass dependence while considering the infrared
region. As it is well known, the ``leading UV logs", which after the RG
summation yield, in particular, the Landau pole, arise from the one-loop
Feynman diagrams. However, in the IR region these diagrams behavior,
generally, is far from being logarithmic. For instance, to the virtual
dissociation of a vector particle (photon, gluon) into a massive
fermion-antifermion pair ($e^++e^-$\,;\, $\,q+\bar{q}\,$) in the $s$--wave
state, there corresponds a function $\,I_s(Q^2/M^2)\,$ that can be
represented in the form of a spectral integral

\begin{eqnarray}
I_s(z)=z\int\limits_{1}^{\infty}\frac{k_s(\sigma)\,d\,\sigma}{\sigma
(\sigma+z)}; \;\; k_s(\sigma)=\sqrt{\frac{\sigma-1}{\sigma}}
\left(1+\frac{1}{2\sigma}\right)
\label{fot-pol}\end{eqnarray}
and in the space-like region $\,z>0\,$ is a positive, monotonically growing
function with logarithmic asymptotic behavior
$\,I_s(z) \simeq \ln z - C_s +O(z^{-1})\,; C_s= 5/3. $
  For definiteness and in accordance with the QED tradition we have subtracted
it at $\,z=0\,$. This agreement turns out to be convenient for our purpose
in the ``low  $\,Q^2\,$ region". \par
  In QCD, besides $\,I_s(z)\,$, essential is a function  $\,I_p(z)\,$
describing a virtual dissociation of a vector gluon into a pair of massive
vector or scalar particles (gluons or ``ghosts") in the $p$--wave state. This
function can also be represented in the form (\ref{fot-pol}) with an adequate
weight function $\,k_p(\sigma)\,$ and obeys the same simple properties at
$\,z>0\,$. In particular, in the UV limit $\,I_p(z)\simeq\ln z-C_p+O(1/z).$
In our analysis only asymptotic constant $C_p$ will be of importance. It
relates to the integral over the phase volume. Due to this $C_p >C_s$. For
the qualitative estimate we use \cite{mikh} $C_p = 8/3$.
\medskip

 Here, we assume that a virtual gluon has a finite (reasonably small --- see
below) mass $\,m_{\rm gl}\,$. We postpone for the future any detailed
discussion of a possible origin of this mass noting that the mechanism should
be based upon a deeper understanding of the ground state structure of the
quantum gauge $\,SU(3)\,$ field. As a provisional {\it ad hoc} working model
one could imply the picture of spontaneous symmetry breaking, analogous to
the $\,SU(2)\,$ case (for a fresh discussion of the subject, see, e.g.,
Ref.\cite{kevin}). In addition, one should have in mind the upper
phenomenological bound

$$ m_{\rm gl} \lesssim 600 \ \mbox{MeV}\, ,$$ 
corresponding to the absence of a direct experimental signal for this mass
existence.

\section{ Massive Renorm--group}

  For an analysis of the invariant QCD coupling at small space-like values of
the variable $\,Q^2<\Lambda^2\,$ we use the ``massive", that is mass
dependent, renormalization group as it has been explicitly formulated
in the pioneer papers \cite{dan55} in the mid-fifties. \par

 In particular, we shall exploit the fact that massive RG, quite in parallel
to the widely used massless one, sums all iterations of a one-loop
contribution in the invariant coupling
$$ a_s(Q^2)_{\rm pert} =a_s-a_s^2
A_1(Q^2, m^2)+a_s^3\left(A_1(Q^2, m^2)\right)^2 +\dots $$
into the geometric progression \cite{blank}
\begin{eqnarray}
a_s(Q^2)_{\rm rg,1} = \frac{a_s}{1+ a_s A_1(Q^2, m^2)} \; .
 \label{a1rg}\end{eqnarray}
Here, we use the notation for the so-called {\it couplant}
$$ a_s(Q^2)=\alpha_s(Q^2)/4\pi\;, \;\; a_s=\alpha_s/4\pi\;.$$

Analogous (approximate) RG summed mass--dependent expressions are known
\cite{mass} for the two-loop case --- see below Eqs. (\ref{a2pert}) and
(\ref{a2rg}) --- as well.

   Let us make a comment on the renormalization (subtraction) scheme (RS). We
use the subtraction at $\,Q^2=0\,$, that is the MOM--scheme instead of the
massless $\,\overline{\rm MS}\,$ one which is in common practice.  Among QCD
practitioners there exists a strong prejudice against any MOM--schemes due to
their gauge dependence. For our analysis it is essential to use
mass-dependent expressions for the diagram contributions. The particular RS
is not of principal importance. One could transit from our scheme to a
$\,\overline{\rm MS}\,$ scheme by standard rules. In particular, connection
between MOM, massive $\,\overline{\rm MS}\,$ and the popular massless
$\,\overline{\rm MS}\,$ schemes has been discussed in detail in papers
\cite{mass,doks-mikh} and we have no possibility of repeating the discusson
in this short note.

\section{The one-loop analysis}

 For the one-loop contribution to an invariant couplant we use the
expression\footnote{ Generally, the $\,A_1(Q^2, m^2)\,$ should combine
several (propagator
and vertex) contributions. However, for our semiquantitive analysis only
UV behaviour with asymptotic constants $C_{s,p}$ will be essential.}
\begin{eqnarray}
A_1(Q^2, m^2) = 11\,I_s\left(\frac{Q^2}{m^2_{\rm gl}}\right)-\frac{2}{3}\,
\sum_qI_p\left(\frac{Q^2} {m_q^2}\right)\, .
\label{A1mass}\end{eqnarray}
This expression by convention turns to zero at $\,Q^2=0\,$.
Hence, in (\ref{a1rg}) $\,a_s=a_s(0)\,$. In the opposite limit at $\, Q^2 \gg
m^2\,$, we have
$$ A_1(Q^2, m^2)\to 11\left(\ln\frac{Q^2}{m^2_{\rm gl}}-\frac{8}{3} \right) -
\frac{2}{3}\sum_{q}\left(\ln\frac{Q^2}{m^2_q} -\frac{5}{3}\right)\,.$$
 In the three-quark region, this gives
 $$ A_1(Q^2, m^2) \simeq 9\,\ln \frac{Q^2}{m^2_{\rm gl}} -
2\,\ln \frac{m^2_{\rm gl}}{m^2_s}-\frac{4}{3}\ln \frac{m_s^2}{m_u m_d} -26 \; ;
\ \  Q^2 \gg m_c^2\,, m^2_{\rm gl}\; . $$

  In the massless case to the expression $1/a +A_1(Q^2, m^2)\,$ there
corresponds $\,9\ln(Q^2/\Lambda^2)\,$. Hence, we get the relation
$$\frac{1}{a_s}+ 22\,\ln \frac{m_s}{m_{\rm gl}}+ 18\ln \frac{\Lambda}{m_s}
= 26 +  \frac{4}{3} \ln \frac{m_s^2}{m_u m_d} $$
between the combination $\,1/a-22\ln m_{\rm gl}\,$ and the QCD scale parameter
$\,\Lambda\,$. For a quantitative estimate let us define an ``effective
one-loop QCD scale parameter" from the condition
$\,\alpha_s(M_{\tau}^2)=0.37\,$ that yields
 $\,\Lambda_1 = 250 \,$ MeV. \par
Now putting $\,m_u= 5\, \mbox{MeV},m_d=10\; \mbox{MeV},\ m_s=150 \;\mbox{MeV}
\,,\, m_{\rm gl}= m_s/\sigma\,$, we arrive at the relation
\begin{eqnarray}
\frac{1}{a_s}  = 25 - 22 \ln \sigma \; ;\ \ \
 {\alpha}_s(0) =\frac{2\pi}{12.5+ 11\ln (m_{\rm gl}/m_s) }\;.
\label{basic}\end{eqnarray}
  The most important qualitative result that follows from it consists in the
existence of a reasonably small lower bound for the gluon mass
corresponding to the $\,\alpha_s(0) \simeq 1\,$ condition.

 We have
$$100 \,\mbox{MeV} < m_{\rm gl} < 600 \ \mbox{MeV} \, .  $$

   In the Table a few values of the IR limit $\,\alpha_s(0)\,$
for some $\,m_{\rm gl}\,$ of this interval are given.
\begin{center}
 {\sf\large Table}    \medskip

\begin{tabular}{|c||c|c|c|c|c|c|}  \hline
  $m_{\rm gl}/$MeV &  100   &  150  & 200  & 300   & 450 \\  \hline
 $\alpha_s(0)$     &  0.87  & 0.55  & 0.41 & 0.36  & 0.31 \\ \hline
\end{tabular}
\end{center}

  As it follows from the Table, there exists a rather wide interval of the
gluonic mass $\,m_{\rm gl}\,$  with ``reasonably small" $\,\alpha_s(0)
\lesssim 1\,$ values.

 Let us note that the lower bound as obtained from relation (\ref{basic}) has
only a qualitative nature. The point is that due to the relation $\,m_{\rm
u,d}\ll m_{\rm gl}\,$ the derivative $ d a(x)/d x\,$ at $\,x=0\,$ is positive
and in the region $\,0<Q^2< m^2_{\rm gl}\,$ the light quark contribution
dominates.  The real place of possible ``blowing up" is close to
$\,Q^2_*=2m^2_{\rm gl}\,$.  As a result, the real lower bound for $m_{\rm
gl}$ corresponds to the condition $ a_s<1/(28-4\ln\sigma)\, ,$ and turns out
to be close to the strange quark mass.

 The upper bound existence for the $\,\alpha_s(0) \,$ value resembles us the
property of unitary models for lower partial waves of hadron scattering in
the low-energy elastic region. These models were popular in the sixties --
see, e.g., Refs.~\cite{tzu61}. Besides analyticity, they satisfy the
two-particle unitarity condition. One-loop diagrams summed in the solution
(\ref{a1rg}), (\ref{A1mass}) just relate to this last condition . This is the
reason for associating our construction with unitarity.

\section{Two--loop corrections}

   For a more accurate numerical description of the invariant coupling in
the IR region one can use the two-loop massive perturbation expansion
\begin{equation}
a_s(Q^2)_{\rm pert,2} =a_s-a_s^2 A_1(Q^2, m^2)+
+ a_s^3A_1^2 - a_s^3\,A_2(Q^2, m^2)+ \dots \, \,.
\label{a2pert}\end{equation}
  An approximate two-loop massive RG solution is of the form \cite{mass}
\begin{eqnarray}
a_s(Q^2)_{\rm rg,2} =a_s\left\{1+a_s A_1(Q^2, m^2)+ a_s\frac{A_2(Q^2, m^2)}
{A_1(Q^2, m^2)}\,\ln \left(1+ a_s A_1(Q^2, m^2) \right)\right\}^{-1}\,.
\label{a2rg}\end{eqnarray}
  At small $a_s$ values this expression corresponds to (\ref{a2pert}).
At the same time, at $\,Q^2 \gg m^2$ it can be represented in the usual form
$$ a^{-1}_s(Q^2)_{\rm rg,2} \to \beta_1\left\{ \ln \frac{Q^2}{\Lambda^2} +
b_1 \ln \left(\ln \frac{Q^2}{\Lambda^2} \right) \right\}\,; \;\; b_1=
\frac{\beta_2}{\beta_1^2}\;.$$

  As it follows from (\ref{a2rg}), in the IR region
\begin{eqnarray}
\frac{1}{a_s(Q^2)} = \frac{1}{a_s} + A_1(Q^2, m^2) +a_s A_2(Q^2, m^2) \;.
\end{eqnarray}

 Hence, the two-loop contribution $\,A_2(Q^2, m^2)\,$ is here suppressed by a
small numerical factor $\,a_s= \alpha_s(0)/4\pi \lesssim 1/10\,$ and cannot
seriously influence the one--loop estimate obtained above.

\section{Conclusion}
  The estimate obtained shows that an accurate description of the threshold
effects related to the light quarks and, especially, to gluons allows us to
formulate one more resolution of the issue of unphysical IR singularities in
the pQCD. The price of this resolving consists in introducing of the only
parameter -- the gluon finite mass.

Thence, it is possible to keep the QCD invariant coupling $\,\alpha_s(Q^2)\,$
in the weak coupling domain in the whole space-like region $\,0<Q^2<\infty\,$
for the gluon mass values
$$ 200 \ \mbox{\rm MeV}< m_{\rm gl}< 600 \ \mbox{\rm MeV}\,.$$
 This theoretical ``window" for the gluonic mass could be enlarged if we
change the light quark masses from their current values to some effective
ones $\,m_{\rm u,d}^{eff} \thicksim m_{\pi}\,$ \cite{jeg}. Here, one can use
the gluon effective mass of the same order of magnitude. \par
\medskip

  To conclude, let us remark that the present model, in reality, is not very
far from our previous construction Refs.\cite{aa96-97} with an explicit
introducing of nonperturbative contribution. A gluon mass $\,m_{\rm gl}$,
being considered in the light of the genuine gauge-invariant QCD Lagrangian,
certainly represents an effective nonperturbative parameter. The same is true
for the abovementioned $\,m_{\rm u,d}^{eff} $.

  In our opinion, this demonstrates once more (compare, e.g., with
discussion of the QED case in Ref.\cite{qed-ghost}) that the ghost-pole
trouble is not a physical problem. It is a technical drawback inherent to
usual, the pQCD one, way of theoretical analysis.

{\section*{Acknowledgements}

  It is a pleasure to thank Drs. D.Yu.\ Bardin, D.I.\ Kazakov, S.V.\
Mikhailov and I.L.\ Solovtsov for the interest in this research and
stimulating discussion. The partial support of the RFFI grants Nos.
96-15-96030, 99-01-00091, and INTAS 96-0842 is appreciated.


\begin{thebibliography}{50}
\bibitem{ms} A.C. Mattingly and P.M. Stevenson, {\it Phys. Rev.}\/ {\bf D 49}
            437 (1994).
\bibitem{sim} Yu.A. Simonov, {\it Yad. Fizika}\/ {\bf 58} (1995) 1139;
        A.M. Badalian and Yu.A. Simonov, {\it ibid.}\/ {\bf 60} (1995) 714.
\bibitem{aa96-97} D.V.~Shirkov and I.L.~Solovtsov,
{\it JINR Rapid Comm.}\/ No. 2[76]-96, 5-10, hep-ph/9604363;
  {\it Phys. Rev. Lett.} \, {\bf 79} (1997) 1209-12, hep-ph/9704333.
\bibitem{tmp99} D.V.~Shirkov, ``Renorm-group, Causality and Non-power
     Perturbation Expansion in QFT", hep-th/9810246; JINR preprint E2-98-311;
     to appear in the April issue {\it Teor. Mat. Fizika}\/ {\bf 119} (1999)
        No.1.
\bibitem{mikh}  S. Mikhailov, private communication.
\bibitem{kevin}  Kevin Cahill, ``Chiral Symmetry and Quark Confinement",
		  hep-ph/9812312.
\bibitem{dan55} N.N. Bogoliubov and D.V.Shirkov, {\it Doklady AN SSSR}\/
	    {\bf 103} (1955) 203--6; 391--4 (in Russian) -- see also
         {\it Sov. Phys. JETP} \/
          {\bf 3} (1956) 77--86; {\it Nuovo Cim.} \/{\bf 3} (1956) 57--64.
\bibitem{blank}  V.Z. Blank  and D.V.Shirkov, {\it Nucl. Phys.} \/{\bf 2},
              (1956/57) 356--70
\bibitem{mass} D.V.~Shirkov, {\it Nucl. Phys.}\/ {\bf B 371} (1992) 467-81;
               {\it Theor. Math. Phys.}\/ {\bf 93} Dec 1992, 466-472; in
           {\sf Perspectives in Particle Physics} \/, Eds. D.Klabucar et al.,
           WS, 1995, pp 1--13; in {\sf Proc. EPSHEP95 Conf.} (Brussels, July
            1995), Eds. J.Lemonne et al., WS, pp 141-2.
\bibitem{doks-mikh} Yu.L.\,Dokshitzer and D.V.~Shirkov, {\it Zeit. Phys.C}
          \/ {\bf 67} (1995) 449-58; \par
           D.V. Shirkov and S.V. Mikhailov, {\it Zeit. Phys.C}\/ {\bf 63}
          (1994) 463-469.
\bibitem{tzu61} A.V. Efremov, D.V. Shirkov, and H.Y.\,Tzu, {\it Sov. Phys.
            JETP}\/ {\bf 14} (1961/1962) 432-7; {\it Scientia Sinica}\/ {\bf
           10} (1961) pp 812-36 -- see also Section 11.3 in the monograph by
         D.V. Shirkov, V.A. Meshcheryakov and V.V. Serebryakov,
             {\sf Dispersion Theories of Strong Interactions at Low
             Energy}, North--Holland, 1969.
\bibitem{pl97} I.L.~Solovtsov and D.V.~Shirkov, {\it Phys. Lett.}\/ {\bf B
               442} (1998) 344-8, also hep-ph/9711251.
\bibitem{qcd97} D.V.~Shirkov, {\it Nucl. Phys. B \/(Proc. Suppl.)}\/ {\bf 64},
          (1998) 106-9, hep-ph/9708480.
\bibitem{jeg} F. Jegerlehner, {\it Nucl. Phys.} Proc.Suppl. {\bf C 51} (1996)
	     131; DESY Preprint 96-121, June 1995; hep-ph/9606484.
\bibitem{qed-ghost} N.N.~Bogoliubov and D.V.~Shirkov, {\it Doklady AN SSSR}
         {\bf 105} (1955) 685-688 (in Russian); see also Section ``Asymptotic
        behavior in the UV" in Chapter ``Renormalization group" of the
      monograph by N.N.~Bogoliubov and D.V.~Shirkov, {\sf Introduction to the
      theory of quantized fields}; two American editions 1959 (Section 43.2,
      pp 528-529) and 1980 (Section 50.2, page 518), Wiley-Interscience, N.Y.;
      Bogolioubov N.N., Chirkov D.V.~ {\sf Introduction \`a la Theorie des
      Champes Quantique}, Paris, Dunod, 1960 (Section 43.2, pp 436-437).
\end{thebibliography}
\end{document}